\documentclass[conference]{IEEEtran}
\IEEEoverridecommandlockouts
\usepackage{cite}
\usepackage{amsmath,amssymb,amsfonts}
\usepackage{algorithmic}
\usepackage{graphicx}
\usepackage{textcomp}
\usepackage{xcolor}
\usepackage{tikz}
\usepackage{dsfont}
\usetikzlibrary{positioning}
\usepackage{mathtools,amssymb,lipsum}

\usepackage{cuted}
\usepackage{multirow}
\def\BibTeX{{\rm B\kern-.05em{\sc i\kern-.025em b}\kern-.08em
    T\kern-.1667em\lower.7ex\hbox{E}\kern-.125emX}}
\begin{document}

\title{MIMO Detection via Gaussian Mixture Expectation Propagation: A Bayesian Machine Learning Approach for High-Order High-Dimensional MIMO Systems\\
}

\author{\IEEEauthorblockN{1\textsuperscript{st} Shachar Shayovitz}
\IEEEauthorblockA{
\textit{Toga Networks}\\
Hod Hasharon, Israel \\
shachar.shayovitz@huawei.com}
\and
\IEEEauthorblockN{2\textsuperscript{nd} Doron Ezri}
\IEEEauthorblockA{
\textit{Toga Networks}\\
Hod Hasharon, Israel \\
doron.ezri@huawei.com}
\and
\IEEEauthorblockN{3\textsuperscript{rd} Yoav Levinbook}
\IEEEauthorblockA{\textit{Toga Networks}\\
Hod Hasharon, Israel \\
yoav.levinbook@huawei.com}
}

\maketitle

\begin{abstract}
MIMO systems can simultaneously transmit multiple data streams within the same frequency band, thus exploiting the spatial dimension to enhance performance. MIMO detection poses considerable challenges due to the interference and noise introduced by the concurrent transmission of multiple streams. Efficient Uplink (UL) MIMO detection algorithms are crucial for decoding these signals accurately and ensuring robust communication. In this paper a MIMO detection algorithm is proposed which improves over the Expectation Propagation (EP) algorithm. The proposed algorithm is based on a Gaussian Mixture Model (GMM) approximation for Belief Propagation (BP) and EP messages. The GMM messages better approximate the data prior when EP fails to do so and thus improve detection. This algorithm outperforms state of the art detection algorithms while maintaining low computational complexity.

\end{abstract}


\section{Introduction}
In recent years, Multiple-Input Multiple-Output (MIMO) technology has emerged as a cornerstone in modern wireless communication systems, offering significant improvements in spectral efficiency, reliability, and capacity \cite{goldsmith2003capacity,zheng2003diversity}. By leveraging multiple antennas at both the transmitter and receiver, MIMO systems can simultaneously transmit multiple data streams within the same frequency band, thus exploiting the spatial dimension to enhance performance.

The detection of signals in MIMO systems, however, poses considerable challenges due to the interference and noise introduced by the concurrent transmission of multiple streams. This task is increasingly challenging as the dimensions of the system (number of transmit and receive antennas) and the constellation order (the number of symbols in the modulation scheme) grow. 

Traditional detection methods, such as Maximum Likelihood Detection (MLD), provide optimal performance but are computationally prohibitive for large-scale MIMO systems. The complexity of MLD grows exponentially with the number of antennas and the constellation order. For instance, in a system with \(N_t\) transmit antennas and \(M\)-QAM modulation, the ML detector needs to evaluate \(M^{N_t}\) possible transmitted symbol combinations. This exponential growth makes MLD infeasible for large MIMO systems, especially in real-time applications. As a result, suboptimal yet computationally efficient algorithms, including Linear Detectors (e.g., Zero Forcing, Minimum Mean Squared Error), Successive Interference Cancellation, and various heuristic and probabilistic approaches, have been extensively studied \cite{he2020model,boutros2003soft,burg2005vlsi,goldberger2011mimo,goldberger2013improved,guo2006algorithm,liu2008modified,studer2008soft,vsvavc2013soft}.

Message-passing algorithms have gained prominence in MIMO detection due to their ability to efficiently handle the inherent complexity of these systems. These algorithms, inspired by Belief Propagation (BP) \cite{yedidia2003understanding} in graphical models, iteratively exchange messages between nodes in a Factor Graph (FG) representing the joint posterior probability distribution of the transmitted symbols given received observations. This iterative approach facilitates the computation of marginal posterior distribution over the transmitted symbols, enabling effective detection and decoding.

Among the various message-passing techniques, the Sum-Product Algorithm (SPA) and its variants, such as the Approximate Message Passing (AMP) and the Generalized Belief Propagation (GBP), have been extensively studied. These algorithms \cite{cespedes2014expectation,jeon2015optimality,wu2014low} leverage the structure of the MIMO channel model to simplify the detection process, achieving a balance between computational complexity and detection accuracy.

Expectation Propagation (EP) \cite{minka2013expectation}, a Bayesian Machine Learning technique which is a more recent development in the field, has also shown promise in MIMO detection. EP extends traditional message-passing algorithms by approximating the posterior distribution with a series of moment matching steps, which can improve the accuracy of the approximations. This technique has been successfully applied to various problems in signal processing and machine learning, demonstrating its potential in enhancing the performance of MIMO detection systems. 

In summary, the combination of exponential complexity, high-dimensional search space, amplified noise and interference effects, and the need for scalable algorithms make MIMO detection in large dimensions with high constellation orders a formidable challenge. This motivates the ongoing research into more efficient and robust detection algorithms that can operate effectively in these challenging scenarios.

In this paper, a MIMO detection algorithm based on a GMM approximation for EP is proposed. It is observed that the true prior for the data symbols is a discrete uniform distribution and thus the Gaussian prior used in Linear Minimum Mean Square Error (LMMSE) and EP is not accurate. We propose to approximate certain EP messages as GMMs and improve the resulting posterior accuracy.

\section{System Model}
In this section, the mathematical model for the received UL signal is defined. Let $n$ be the number of data symbols from some constellation (for simplicity we assume the same constellation for all symbols), where $\mathcal{A}$ denotes the set of symbols in the constellation and $E_s$ is the mean symbol energy. The transmitted symbol vector $\mathbf{u} \in \mathcal{A}^{n}$ is an $n \times 1$ i.i.d. vector. The symbols are transmitted over a flat-fading complex MIMO channel defined by $\mathbf{H} \in \mathbb{C}^{m \times n}$, where each coefficient is drawn according to a proper complex zero-mean unit-variance Gaussian distribution and $m$ is the number of receiving antennas. 

The channel output $\mathbf{y} \in \mathbb{C}^{m}$ is given by:
\begin{equation}\label{eq:system_model}
    \mathbf{y}=\mathbf{H}\mathbf{u}+\mathbf{n}
\end{equation}
where $\mathbf{n}$ is an additive white circular-symmetric complex Gaussian noise vector with independent zero-mean components and $\sigma^2_w$ variance.

Given the model in (\ref{eq:system_model}), the posterior probability for the transmitted symbol vector $\mathbf{u}$:
\begin{equation}\label{eq:prob_model}
    p\left(\mathbf{u}|\mathbf{y}\right)\propto p\left(\mathbf{y}|\mathbf{u}\right)p\left(\mathbf{u}\right) \propto \mathcal{N}(\mathbf{y};\mathbf{H}\mathbf{u},\,\sigma_w^{2}\mathbf{I})\Pi_{i=1}^n \mathds{1}_{u_i \in \mathcal{A}}
\end{equation}
where $\mathds{1}_{u_i \in \mathcal{A}}$ is the indicator function that takes value one
if $u_i \in  \mathcal{A}$ and zero otherwise. Note that $p(\mathbf{u}) \propto \Pi_{i=1}^n \mathds{1}_{u_i \in \mathcal{A}}$ is uniform across all points in $\mathcal{A}^{n}$.

Similarly to the formulation in \cite{cespedes2014expectation}, the MIMO model is transformed from complex valued to real valued by an equivalent double-sized real-valued representation. This representation is obtained by considering the real and imaginary parts separately. The complex notations are the same as the real notations to the keep the notations uncluttered.

The minimum symbol error optimal detection rule is the Maximum A-Posteriori (MAP) detector which requires the knowledge of the posterior distribution $p\left(u_i|\mathbf{y}\right)$. This posterior can be calculated using marginalization on the joint posterior proposed in (\ref{eq:prob_model}):
\begin{equation}\label{eq:marginal_pu}
    p\left(u_i|\mathbf{y}\right)\propto \int_{\mathbf{u}^{-i}}\mathcal{N}(\mathbf{y};\mathbf{H}\mathbf{u},\sigma_w^{2}\mathbf{I})\Pi_{j=1}^N \mathds{1}_{u_j \in \mathcal{A}} \,d\mathbf{u}^{-i}
\end{equation}
where $\mathbf{u}^{-i} = [u_1,u_2,...,u_{i-1},u_{i+1},...,u_m]$ is the vector of all symbols except $u_i$.

The multi-dimensional integral in (\ref{eq:marginal_pu}) is intractable and approximations are needed in order to find a solution. This is exactly where the BP algorithm comes in and provides an iterative message passing algorithm which computes an approximation of this integral for all $u_i$. Since this work is focused on the MAP detector for the marginal posterior, we will use the Sum Product Algorithm \cite{loeliger2007factor} which is a specific variant of the BP algorithm for computing marginal posteriors.  

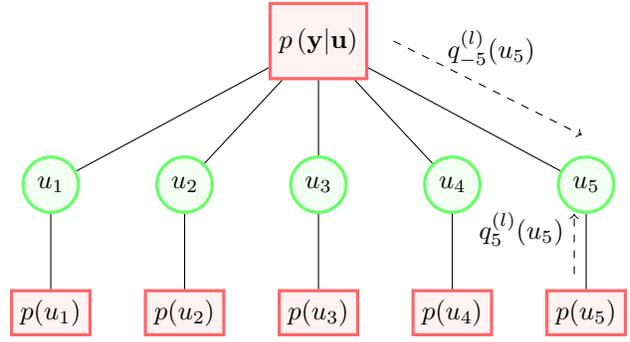
\begin{figure}
\begin{tikzpicture}[
roundnode/.style={circle, draw=green!60, fill=green!5, very thick, minimum size=5mm},
squarednode_main/.style={rectangle, draw=red!60, fill=red!5, very thick, minimum size=10mm},
squarednode/.style={rectangle, draw=red!60, fill=red!5, very thick, minimum size=5mm},
]
\node[squarednode_main]      (channel)                              {$p\left(\mathbf{y}|\mathbf{u}\right)$};
\node[roundnode]        (u3)       [below=of channel] {$u_3$};
\node[roundnode]        (u2)       [below=of channel, left=of u3] {$u_2$};
\node[roundnode]        (u1)       [below=of channel, left=of u2] {$u_1$};
\node[roundnode]        (u4)       [below=of channel, right=of u3] {$u_4$};
\node[roundnode]        (u5)       [below=of channel, right=of u4] {$u_5$};
\node[squarednode]      (p_u1)       [below=of u1] {$p(u_1)$};
\node[squarednode]      (p_u2)       [below=of u2] {$p(u_2)$};
\node[squarednode]      (p_u3)       [below=of u3] {$p(u_3)$};
\node[squarednode]      (p_u4)       [below=of u4] {$p(u_4)$};
\node[squarednode]      (p_u5)       [below=of u5] {$p(u_5)$};

\draw[-] (channel) -- (u1);
\draw[-] (channel) -- (u2);
\draw[-] (channel) -- (u3);
\draw[-] (channel) -- (u4);
\draw[-] (channel) -- (u5);
\draw[-] (u1.south) -- (p_u1.north);
\draw[-] (u2.south) -- (p_u2.north);
\draw[-] (u3.south) -- (p_u3.north);
\draw[-] (u4.south) -- (p_u4.north);
\draw[-] (u5.south) -- (p_u5.north);
\draw[dashed, ->] (3.4,-3.1) -- (3.4,-2.3);
\node at (2.7,-2.5) {$q^{(l)}_{5}(u_5)$};
\draw[dashed, ->] (1,0) -- (3.5,-1.25);
\node at (2.3,-0.1) {$q^{(l)}_{-5}(u_5)$};
\end{tikzpicture}
\caption{Factor Graph Representation for 5 data streams MIMO Channel}
\label{fig:fg}
\end{figure}

In order to define the SPA messages, we first introduce the reader to Factor Graphs. A Factor Graph (FG) is a bipartite graph representing the factorization of a function. In probability theory and its applications, FGs are used to represent factorization of a probability distribution function, enabling efficient computations, such as the computation of marginal distributions through the sum–product algorithm. The FG for the posterior in (\ref{eq:prob_model}) is shown in Fig.\ref{fig:fg} for $m=5$. The SPA messages are defined in \cite{loeliger2007factor} for a general FG.

\section{MIMO Detection via Expectation Propagation}
A straightforward implementation of SPA for the FG described Fig.\ref{fig:fg} is still too complex since the distribution $\mathcal{N}(\mathbf{y};\mathbf{H}\mathbf{u},\,\sigma_w^{2}\mathbf{I})$ does not factor. In \cite{cespedes2014expectation}, Expectation Propagation (EP) \cite{minka2013expectation} is used in order to perform low complexity MIMO detection for the same FG. EP is an approximation of the BP algorithm \cite{yedidia2003understanding}. It is an iterative message passing algorithm which estimates the posterior on each data stream $u_i$ while improving the posterior with every iteration. The EP messages are Gaussians which propagate through the FG and iteratively refine the posterior of each data symbol. Effectively the mean and variance are the messages being passed between the nodes in the FG. In this section, for mathematical completeness, we will detail the messages of the EP MIMO detection algorithm as provided in \cite{cespedes2014expectation}.

\subsection{Cavity Update}
For all iterations $l \geq 0$, EP uses Gaussian priors for the messages $q^{(l)}_{i}\left(u_{i}\right)$ and updates their parameters (mean and covariance) every iteration:
\begin{equation}\label{eq:prior_update}
    q^{(l)}_{i}\left(u_{i}\right) = \mathcal{N}(u_{i};\gamma^{(l)}_{i}\Lambda_{i}^{{-1}^{(l)}},\Lambda_{i}^{{-1}^{(l)}})
\end{equation}

The Gaussian priors propagate upwards in the FG as can be seen for $i=5$ in Fig.\ref{fig:fg}. All these messages (priors) are multiplied at the function node and the joint posterior is updated:
\begin{equation}
    q^{(l)}\left(\mathbf{u}|\mathbf{y}\right)\propto \mathcal{N}(\mathbf{y};\mathbf{H}\mathbf{u},\sigma_w^{2}\mathbf{I})\Pi_{i=1}^N q^{(l)}_i\left(u_i\right)
\end{equation}
Multiplication of Gaussians produces a Gaussian with the following mean and covariance:
\begin{equation}\label{eq:cov_mmse}
    \mathbf{\Sigma^{(l)}} = \left(\sigma_w^{-2}\mathbf{H}^T\mathbf{H}+\text{diag}\left(\mathbf{\Lambda}^{(l)}\right)\right)^{-1} 
\end{equation}

\begin{equation}\label{eq:mean_mmse}
    \boldsymbol{\mu}^{(l)} = \mathbf{\Sigma}^{(l)}\left(\sigma_w^{-2}\mathbf{H}^T\mathbf{y}+\boldsymbol{\gamma}^{(l)}\right)
\end{equation}
The resulting Gaussian mean (\ref{eq:mean_mmse}) is effectively the LMMSE estimator based on the updated Gaussian priors. For $l=0$, the priors are zero mean Gaussians and therefore we get the conventional MIMO LMMSE detector.

The messages propagating downwards in the FG (denoted cavity in \cite{cespedes2014expectation}) are computed based on the BP algorithm:
\begin{equation}\label{eq:cavity_gauss}
    q_{-i}^{(l)}\left(u_i|\mathbf{y}\right)\propto \int_{\mathbf{u}^{-i}}\mathcal{N}(\mathbf{y};\mathbf{H}\mathbf{u},\sigma_w^{2}\mathbf{I})\Pi_{j \neq i} q^{(l)}_j\left(u_j\right)\,d\mathbf{u}^{-i}
\end{equation}
Since the priors are Gaussians, then (\ref{eq:cavity_gauss}) is a Gaussian with a simple analytical expression. The resulting Gaussian takes the corresponding elements from the mean vector, $\mu^{(l)}_i$, in (\ref{eq:mean_mmse}) and covariance matrix diagonal, $\sigma_i^{{2}^{(l)}}$, in (\ref{eq:cov_mmse}):
\begin{equation}\label{eq:cavity_gauss_1}
    q_{-i}^{(l)}\left(u_i|\mathbf{y}\right) =  \mathcal{N}(u_i;t^{(l)}_i,h_i^{{2}^{(l)}})
\end{equation}
where:
\begin{equation*}
    h^{2(l)}_i = \frac{\sigma^{2(l)}_i}{1-\sigma^{2(l)}_i\Lambda_i^{(l)}}
\end{equation*}
\begin{equation*}
    t^{(l)}_i = h^{2(l)}_i\left(\frac{\mu_i^{(l)}}{\sigma^{2(l)}_i}-\gamma^{(l)}_i\right)
\end{equation*}
The $i$-th cavity takes into account the channel model and the information from all the priors associated with data symbols other than the $i$-th symbol. 

In this paper we consider, for the ease of presentation, the uncoded case and an extension to the coded case is straightforward. The cavity messages are used to compute Log Likelihood Ratios (LLRs) which are sent to an error correcting code such as LDPC (Turbo decoding \cite{uchoa2015iterative}).

\subsection{Prior Update}
The main innovation in \cite{cespedes2014expectation} is the method by which the prior, $q^{(l+1)}_i\left(u_i\right)$, is updated. In order to keep $q^{(l+1)}_i\left(u_i\right)$ Gaussian, the following approximation is done on the joint posterior of $u_i$: 
\begin{equation}\label{eq:main_ep_approx}
    q^{(l+1)}_i\left(u_i\right)q^{(l)}_{-i}\left(u_i\right) \approx q^{(l)}_{-i}\left(u_i\right)\mathds{1}_{u_i \in \mathcal{A}}
\end{equation}
The right hand side of (\ref{eq:main_ep_approx}) is not a Gaussian and thus a projection to the Gaussian parametric family needs to be taken (moment matching minimizes the Kullback Liebler Divergence (KLD)). 
\begin{equation}\label{eq:kld_approx}
    \mathcal{N}(u_i;\mu^{(l)}_{p_i},\sigma^{{2}^{(l)}}_{p_i}) = \text{proj}\left( q^{(l)}_{-i}\left(u_i\right)\mathds{1}_{u_i \in \mathcal{A}}\right)
\end{equation}
where $\mu^{(l)}_{p_i}$ and $\sigma^{{2}^{(l)}}_{p_{i}}$ are the mean variance generated using moment matching.

The updated prior is then computed: 
\begin{equation}\label{eq:kld_approx_gmm}
    q^{(l+1)}_i\left(u_i\right) = \frac{\text{proj}\left( q^{(l)}_{-i}\left(u_i\right)\mathds{1}_{u_i \in \mathcal{A}}\right)}{q^{(l)}_{-i}\left(u_i\right)}
\end{equation}

Therefore:
\begin{equation}\label{eq:cavity_ep}
\begin{split}
    &q^{(l+1)}_i\left(u_i\right) = \frac{\mathcal{N}(u_i;\mu^{(l)}_{p_i},\sigma^{{2}^{(l)}}_{p_i})}{\mathcal{N}(u_i;t^{(l)}_i,h_i^{{2}^{(l)}})}
    \\ & \propto \mathcal{N}(u_i;\gamma^{(l+1)}_i\Lambda_i^{{-1}^{(l+1)}},\Lambda_i^{{-1}^{(l+1)}})
    \end{split}
\end{equation}

where:
$$
\gamma^{(l+1)}_i = \frac{\mu_{p_i}^{(l)}}{\sigma^{{2}^{(l)}}_{p_{i}}}-\frac{t_i^{(l)}}{h^{2(l)}_i}
$$

$$
\Lambda_{i}^{(l+1)} = \frac{1}{\sigma^{{2}^{(l)}}_{p_{i}}}-\frac{1}{h^{2(l)}_i}
$$
The process repeats with the updated prior using (\ref{eq:prior_update}).

\section{Gaussian Mixture Messages}
However, the Gaussian division in (\ref{eq:cavity_ep}) can produce a Gaussian with a negative variance when $\sigma^{{2}^{(l)}}_{p_{i}} > h^{2(l)}_i$, meaning that the updated prior, $q^{(l+1)}_i\left(u_i\right)$, is not a normalized Gaussian. In \cite{cespedes2014expectation}, this is mitigated by identifying the negative variance prior messages and replacing them with zero mean and $E_s$ variance Gaussians. Effectively, it means that the updated prior is uninformative and does not utilize the cavity information for the subsequent iteration. 

\begin{figure}
    \centering
    \includegraphics[width=1.0\linewidth]{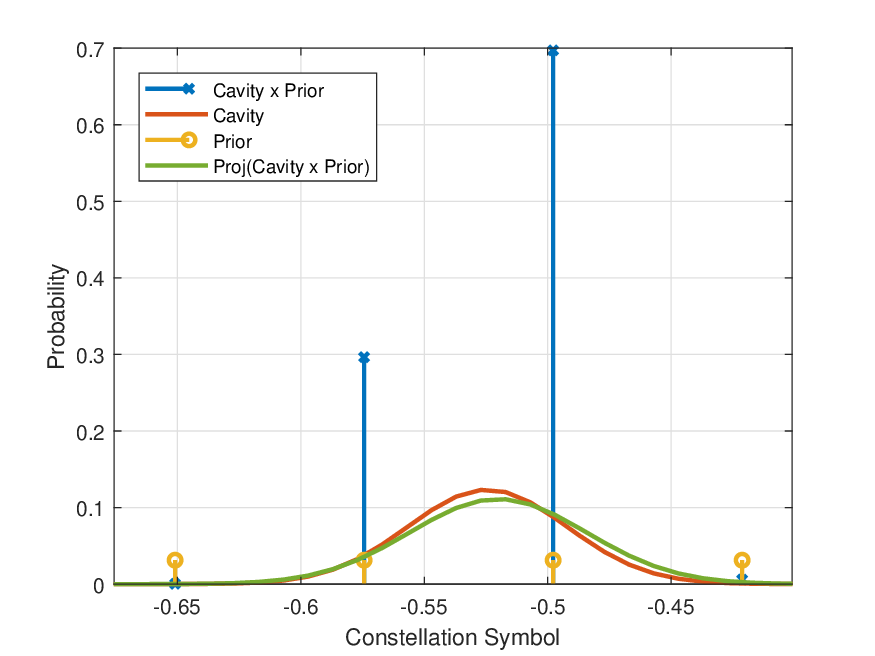}
    \caption{Gaussian Approximation for Messages}
    \label{fig:enter-label}
\end{figure}

In this section, we propose to use a GMM and not a single Gaussian when negative variances arise. We first realize that if only one constellation symbol has a dominant probability in the joint posterior, then $\sigma^{{2}^{(l)}}_{p_{i}} \leq h^{2(l)}_i$. Therefore, when $\sigma^{{2}^{(l)}}_{p_{i}} > h^{2(l)}_i$, there may be several constellation symbols which are similarly likely. An illustration of this is provided in Fig.\ref{fig:enter-label} where we can see the true prior (delta functions), cavity message, joint posterior and the Gaussian approximation for the joint posterior. It can be observed that in this case the variance of the true posterior is larger than the cavity ($\sigma^{{2}^{(l)}}_{p_{i}} > h^{2(l)}_i$), which results with a negative variance for the updated prior. As observed, the multiplication of the prior and cavity distributions differs from a single Gaussian distribution if the variance (uncertainty) of the cavity is larger than the distance between two adjacent real constellation points. Our approach is to improve this approximation so the updated prior will provide refined information to the computation of the updated cavity.

We propose the following approximation for the true prior:
\begin{equation}\label{eq:mixture_approx_indicator}
     \mathds{1}_{u_i \in \mathcal{A}} \approx \sum_{k=1}^K\alpha_k\mathcal{N}\left(u_i;a_k,\sigma^{2}_0\right)
\end{equation}
where $\sigma^{2}_0, a_k \in \mathcal{A}$ and $\alpha_k$ are the variance, mean of a Gaussian around a real value constellation point and the mixing coefficient respectively. This approximation is accurate when $\sigma^{2}_0 \rightarrow 0 $ and $\alpha_k=1$ for all $k$.

Using the mixture approximation (\ref{eq:mixture_approx_indicator}), we can write the prior message for $u_s$:

\begin{equation}
\begin{split}
   q^{(l+1)}_s\left(u_s\right) = \sum_{k=1}^K\alpha_k q^{(l+1)}_{s,k}\left(u_s\right)
\end{split}
\end{equation}
where: 
\begin{equation}
\begin{split}
   q^{(l+1)}_{s,k}\left(u_s\right) = \mathcal{N}(u_s;\gamma^{(l+1)}_{s,k}\Lambda_{s,k}^{{-1}^{(l+1)}},\Lambda_{s,k}^{{-1}^{(l+1)}})
\end{split}
\end{equation}
and $a_k = \gamma^{(l+1)}_{s,k}\Lambda_{s,k}^{{-1}^{(l+1)}}$ and $\sigma^2_0 = \Lambda_{s,k}^{{-1}^{(l+1)}}$.

\subsection{Selection of Variable Node for GMM Approximation}
Ideally we would like to model \textit{all} the variable nodes with negative variance with GMMs but this will incur an unreasonable computational burden. The subsequent cavity computation will include an integral on a multiplication of several mixtures and the complexity becomes exponential. Taking into account complexity considerations, the mixture approximation may be used for a limited number of variable nodes with negative updated prior variance.

\begin{figure}
    \centering
    \includegraphics[width=1.0\linewidth]{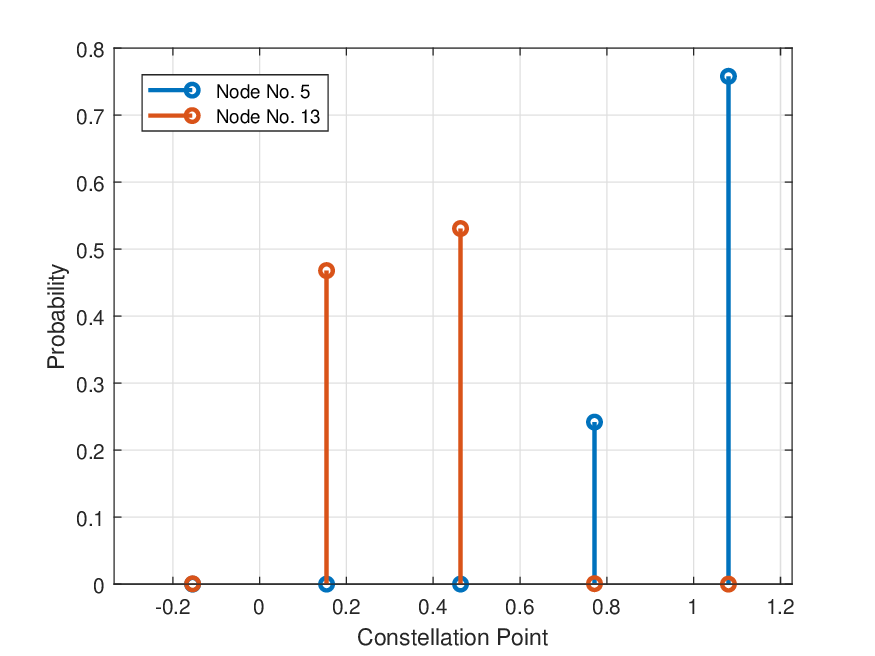}
    \caption{Joint Posterior for Negative Variance Variable Nodes}
    \label{fig:min_entropy}
\end{figure}

In this work, we chose the nodes (corresponding to negative variance) with the lowest entropy of their joint posterior, $q^{(l)}_{-i}\left(u_i\right)\mathds{1}_{u_i \in \mathcal{A}}$. The idea is that these nodes have the lowest uncertainty but their respective EP prior approximation is very uncertain. Therefore, if we model them using the true prior (mixture) then we can infer them more easily than higher entropy posteriors. In order to illustrate this idea, an example from the EP process is provided, where the variable nodes $5$ and $13$ have a negative variance. In Fig.\ref{fig:min_entropy} the joint posteriors for both variable nodes are plotted. The EP algorithm \cite{cespedes2014expectation} will approximate both of them using the same Gaussian. If we can only model one of them using a GMM, we argue that selecting node $5$ would be better. This is since the updated prior will provide better information for the subsequent EP process. This approach was compared to random selection empirically and provided superior results.

The nodes $u_s$ which are selected for mixture message approximation can be changed per message passing iteration. We note also that the prior update for the other nodes is unchanged from the EP algorithm. 

\subsection{Cavity Update with a Gaussian Mixture}
As described in the previous section, the Gaussian priors (and mixture) propagate upwards in the FG as can be seen for $i=5$ in Fig.\ref{fig:fg}. All these messages are multiplied at the function node and the joint posterior is updated. In the following derivation, we consider a single GMM message in the FG for $u_s$. The extension to more GMM messages is straightforward. Based on the BP messages, the cavity for all the nodes \textit{except} $u_s$ is computed using the following expression:
\begin{equation}\label{eq:cavity_with_mix}
\begin{split}
    & q^{(l+1)}_{-i}\left(u_i\right) = \\ &\sum_{k=1}^K\alpha_k\int_{\mathbf{u}^{-i}}\mathcal{N}(\mathbf{y};\mathbf{H}\mathbf{u},\sigma_w^{2}\mathbf{I})\Pi_{j\neq s,i} q^{(l+1)}_j\left(u_j\right) q^{(l+1)}_{s,k}\left(u_s\right)d\mathbf{u}^{-i}
\end{split}
\end{equation}

We can re-write (\ref{eq:cavity_with_mix}) as:
\begin{equation}\label{eq:cavity_with_mix1}
\begin{split}
    q^{(l+1)}_{-i}\left(u_i\right) = \sum_{k=1}^K\alpha_k q_k\left(\mathbf{y}\right)\int_{\mathbf{u}^{-i}}\frac{q^{(l+1)}_k\left(\mathbf{u}|\mathbf{y}\right)}{q^{(l+1)}_i\left(u_i\right)}\,d\mathbf{u}^{-i}
\end{split}
\end{equation}
where:
\begin{equation}\label{eq:mixture_mmse_joint}
\begin{split}
    q^{(l+1)}_k\left(\mathbf{u}|\mathbf{y}\right) = \frac{\mathcal{N}(\mathbf{y};\mathbf{H}\mathbf{u},\sigma_w^{2}\mathbf{I})\Pi_{j\neq s} q^{(l+1)}_j\left(u_j\right) q^{(l+1)}_{s,k}\left(u_s\right)}{q_k\left(\mathbf{y}\right)}
\end{split}
\end{equation}
and 
\begin{equation}\label{eq:mixture_mmse_joint1}
\begin{split}
    q_k\left(\mathbf{y}\right) = \int_{\mathbf{u}}\mathcal{N}(\mathbf{y};\mathbf{H}\mathbf{u},\sigma_w^{2}\mathbf{I})\Pi_{j\neq s} q^{(l+1)}_j\left(u_j\right) q^{(l+1)}_{s,k}\left(u_s\right)\,d\mathbf{u}
\end{split}
\end{equation}
The cavity (\ref{eq:cavity_with_mix1}) is a GMM where the mixing coefficients are $\alpha_k q_k\left(\mathbf{y}\right)$ and $ q_k\left(\mathbf{y}\right)$ are the likelihoods for the observed data, $\mathbf{y}$, based on each mixture component. We observe that (\ref{eq:mixture_mmse_joint}) is a normalized Gaussian with the following mean and covariance:
\begin{equation}\label{eq:mixture_mmse_joint12}
\mathbf{\Sigma}_k^{(l+1)}=\left(\sigma_w^{-2}\mathbf{H}^T\mathbf{H}+\text{diag}\left(\mathbf{\Lambda}_k^{(l+1)}\right)\right)^{-1}
\end{equation}
\begin{equation}\label{eq:mixture_mmse_joint13}
\mathbf{\mu_k^{(l+1)}}=\mathbf{\Sigma}_k^{(l+1)}\left(\sigma_w^{-2}\mathbf{H}^T\mathbf{y}+\mathbf{\gamma}_k^{(l+1)}\right)
\end{equation}
where $\mathbf{\Lambda}_{s,k}^{{-1}^{(l+1)}}=\sigma^2_0$ which means that $\mathbf{\Sigma}_k^{(l+1)}$ is the same for all mixture components, $k$, and thus can be computed once. The mean $\mathbf{\mu}_k^{(l+1)}$ changes for different constellation points (different $\mathbf{\gamma}_k^{(l+1)}$ per $k$). However, its computation does not require matrix inversion and thus is linear with the number mixture components. 

We compute $q_k\left(\mathbf{y}\right)$ using the fact that $\mathbf{y}$ is Gaussian and thus $\mathbf{\mu_y}=\mathbf{H}\mathbf{\mu_u}$ and $\mathbf{\Sigma_y}=\mathbf{H}\mathbf{\Sigma_u}\mathbf{H}^T+\sigma_w^{2}\mathbf{I}$. Therefore:
\begin{equation}\label{eq:likelihood_complex}
    q_k\left(\mathbf{y}\right) \propto e^{-\left(\mathbf{y}-\mathbf{H}\mathbf{\mu_u}\right)^T\left(\mathbf{H}\mathbf{\Lambda}^{(l+1)^{-1}}\mathbf{H}^T+\sigma_w^{2}\mathbf{I}\right)^{-1}\left(\mathbf{y}-\mathbf{H}\mathbf{\mu_u}\right)}
\end{equation}
The reason the Gaussian normalization factor is not taken into account is due to the fact that the covariance, $\mathbf{\Sigma_y}$ is the same for all mixture components. Currently, per EP iteration, the following matrix inversion is performed: $\left(\sigma_w^{-2}\mathbf{H}^T\mathbf{H}+\text{diag}\left(\mathbf{\Lambda}_k^{(l+1)}\right)\right)^{-1}$. However, in (\ref{eq:likelihood_complex}) a different inversion is needed: $\left(\mathbf{H}\mathbf{\Sigma_u}\mathbf{H}^T+\sigma_w^{2}\mathbf{I}\right)^{-1}$. In order to reduce complexity and avoid another direct matrix inversion, we look at the probability of the sufficient statistics, $\left(\mathbf{H}^T\mathbf{H}\right)^{-1}\mathbf{H}^T\mathbf{y}$:
\begin{equation}
    q_k\left(\mathbf{y}\right) \propto e^{-\mathbf{e}^T\left(\mathbf{\Sigma_u}+\sigma_w^{2}\left(\mathbf{H}^T\mathbf{H}\right)^{-1}\right)^{-1}\mathbf{e}}
\end{equation}
where:
\begin{equation}
    \mathbf{e} = \left(\mathbf{H}^T\mathbf{H}\right)^{-1}\mathbf{H}^T\mathbf{y}-\mathbf{\mu_u}
\end{equation}
Using the Matrix Inversion Lemma (MIL) \cite{woodbury1950inverting}:
\begin{equation}
\begin{split}
&\left(\mathbf{\Sigma_u}+\sigma_w^{2}\left(\mathbf{H}^T\mathbf{H}\right)^{-1}\right)^{-1} = \text{diag}\left(\mathbf{\Lambda}_k^{(l+1)}\right) - \\ & \text{diag}\left(\mathbf{\Lambda}_k^{(l+1)}\right)\mathbf{\Sigma}_k^{(l+1)}\text{diag}\left(\mathbf{\Lambda}_k^{(l+1)}\right)
\end{split}
\end{equation}
which uses the inverted matrix already computed and is a multiplication of diagonal matrices which can be further simplified.

Finally, the updated cavity $q^{(l+1)}_{-i}\left(u_i\right)$ is approximated using a Gaussian and thus projection using moment matching of a Gaussian mixture to a single Gaussian is used along with the means and variances from (\ref{eq:mixture_mmse_joint12}) and (\ref{eq:mixture_mmse_joint13}): 
\begin{equation}
\begin{split}
    & q^{(l+1)}_{-i}\left(u_i\right) = \\ &\text{proj}\left(\sum_{k=1}^K\alpha_k q_k\left(\mathbf{y}\right)\frac{\mathcal{N}(u_i;\mu^{(l+1)}_{i,k},\sigma_{i,k}^{{2}^{(l+1)}})}{\mathcal{N}(u_i;\gamma^{(l+1)}_{i}\Lambda_{i}^{{-1}^{(l+1)}},\Lambda_{i}^{{-1}^{(l+1)}})}\right)
\end{split}
\end{equation}

We can re-write,
\begin{equation}
    q^{(l+1)}_{-i}\left(u_i\right) = \text{proj}\left(\sum_{k=1}^K\alpha_k q_k\left(\mathbf{y}\right)\mathcal{N}\left(y;t_{i,k}^{(l+1)},h^{{2}^{(l+1)}}_{i,k}\right)\right)
\end{equation}
where:
\begin{equation*}
    h^{{2}^{(l+1)}}_{i,k} = \frac{\sigma_{i,k}^{{2}^{(l+1)}}}{1-\sigma_{i,k}^{{2}^{(l+1)}}\Lambda_i^{(l+1)}}
\end{equation*}
\begin{equation*}
    t_{i,k}^{(l+1)} = h^{{2}^{(l+1)}}_{i,k}\left(\frac{\mu^{(l+1)}_{i,k}}{\sigma_{i,k}^{{2}^{(l+1)}}}-\gamma^{(l+1)}_i\right)
\end{equation*}

Using moment matching (minimal Kullback Liebler Divergence (KLD)):
\begin{equation}
    q^{(l+1)}_{-i}\left(u_i\right) = \mathcal{N}\left(y;t_{i}^{(l+1)},h^{{2}^{(l+1)}}_{i}\right)
\end{equation}

where:
\begin{equation*}
    t_{i}^{(l+1)} = \sum_{k=1}^K\alpha_k q_k\left(\mathbf{y}\right)t_{i,k}^{(l+1)}
\end{equation*}
\begin{equation*}
    h^{{2}^{(l+1)}}_{i} =  \sum_{k=1}^K\alpha_k q_k\left(\mathbf{y}\right)\left(h^{{2}^{(l+1)}}_{i,k} + \left(t_{i,k}^{(l+1)}-t_{i}^{(l+1)}\right)^2\right)
\end{equation*}
Note that $h^{{2}^{(l+1)}}_{i,k}$ is fixed for all $k$ and that can be used to reduce complexity. 

In order to compute the cavity for $u_s$, the following BP message is used:
\begin{strip}
  \begin{align*}
    q^{(l+1)}_{-s}\left(u_s\right) & =
    \int_{\mathbf{u}^{-s}}\mathcal{N}(\mathbf{y};\mathbf{H}\mathbf{u},\,\sigma_w^{2}I)\Pi_{i \neq s} \mathcal{N}(u_i;\gamma^{(l+1)}_{i}\Lambda_{i}^{{-1}^{(l+1)}},\Lambda_{i}^{{-1}^{(l+1)}})\,d\mathbf{u}^{-s}
    \\
      & = \frac{\mathcal{N}(u_s;\gamma^{(l+1)}_{s,0}\Lambda_{s,0}^{{-1}^{(l+1)}},\Lambda_{s,0}^{{-1}^{(l+1)}})\int_{\mathbf{u}^{-s}}\mathcal{N}(\mathbf{y};\mathbf{H}\mathbf{u},\sigma_w^{2}I)\Pi_{i \neq s} \mathcal{N}(u_i;\gamma^{(l+1)}_{i}\Lambda_{i}^{{-1}^{(l+1)}},\Lambda_{i}^{{-1}^{(l+1)}})d\mathbf{u}^{-s}}{\mathcal{N}(u_s;\gamma^{(l+1)}_{s,0}\Lambda_{s,0}^{{-1}^{(l+1)}},\Lambda_{s,0}^{{-1}^{(l+1)}})}
  \end{align*}
\end{strip}
where $\mathcal{N}(u_s;\gamma^{(l+1)}_{s,0}\Lambda_{s,0}^{{-1}^{(l+1)}},\Lambda_{s,0}^{{-1}^{(l+1)}})$ is essentially a dummy prior which is used in order to simplify the computations and use the variances as the diagonal elements in the covariance of a joint posterior. From a computational complexity standpoint we have already computed the covariance and mean associated with this joint posterior (numerator) in (\ref{eq:mixture_mmse_joint12}) and (\ref{eq:mixture_mmse_joint13}).

Therefore,
\begin{equation}
    q^{(l+1)}_{-s}\left(u_s\right) = \mathcal{N}(y;t_s^{(l+1)},h^{{2}^{(l+1)}}_s)
\end{equation}
where:
\begin{equation*}
    h^{2(l+1)}_s = \frac{\sigma_{s,0}^{{2}^{(l+1)}})}{1-\sigma_{s,0}^{{2}^{(l+1)}}\Lambda_{s,0}^{(l+1)}}
\end{equation*}
\begin{equation*}
    t^{(l+1)}_s = h^{2(l+1)}_s\left(\frac{\mu^{(l+1)}_{s,0}}{\sigma_{s,0}^{{2}^{(l+1)}}}-\gamma^{(l+1)}_{s,0}\right)
\end{equation*}
Note that the covariance and mean of the joint posterior for $u_s$ does not require matrix inversion. We will insert the dummy variance $\Lambda_{s,0}^{{-1}^{(l+1)}}$ into the vector $\mathbf{\Lambda}^{{-1}^{(l+1)}}$ and in order to compute the new covariance, apply MIL. Since the matrix $\text{diag}\left(\mathbf{\Lambda}_k^{(l+1)}\right)$ can be viewed as a sum of rank-1 matrices, the change for $u_s$ is on a single rank-1 matrix.

In order to improve the robustness of the algorithm, in \cite{cespedes2014expectation} it is suggested to smooth the parameter update of the updated priors. In our algorithm we propose to also pass the cavity messages through a low pass filter in order to improve robustness:
\begin{equation*}
    h^{2(l+1)}_i = \beta h^{2(l+1)}_i+(1-\beta)h^{2(l)}_i
\end{equation*}
\begin{equation*}
    t^{(l+1)}_i = \beta t^{(l+1)}_i+(1-\beta)t^{(l)}_i
\end{equation*}
where $\beta \in [0,1]$. We also use all the other techniques for numerical stability as described in \cite{cespedes2014expectation}.

\section{Complexity}
In this section, the computational complexity of the proposed algorithm is discussed and compared to other low complexity MIMO detection schemes. The mixture algorithm, denoted GMEP (Gaussian Mixture Expectation Propagation), shares most computational blocks with the EP algorithm \cite{cespedes2014expectation}. The most computationally demanding block is the matrix inversion which is invoked the same number of times in both algorithms $(\mathcal{O}(n^3))$. The most significant added complexity of GMEP on top of EP is the matrix by vector multiplication in (\ref{eq:mixture_mmse_joint13}): $\mathbf{\Sigma}_k^{(l+1)}\mathbf{\gamma}_k^{(l+1)}$, which is of order $\mathcal{O}(n^2M)$, where $M \leq |\mathcal{A}|$ is the number of mixture components and in fact much smaller than $|\mathcal{A}|$ as shown in simulations. All other matrix multiplications are performed once per variable node and do not depend on the mixture order $M$. The comparison is detailed in Table \ref{Tab:complex}, where $L$ is the number of iterations. 

It is clear that GMEP becomes attractive compared to EP only for $M \ll n$. We have also included an order of magnitude for the complexity of GMEP with 2 mixture messages. Due to the multiplication of messages in the cavity computation, there are $M^2$ multiplication per node in this case.

\begin{table}[ht]
\caption{Computational Complexity}
    \centering
\begin{tabular}{ |p{3cm}||p{5cm}| }
 \hline
 Algorithm& Computational Complexity\\
 \hline
 LMMSE   & $\mathcal{O}(n^3 )$\\
 EP \cite{cespedes2014expectation} &   $\mathcal{O}(n^3L + n|\mathcal{A}|L )$\\
 GMEP (1 Node Mix) & $\mathcal{O}(n^3L + n|\mathcal{A}|L + n^2ML )$\\
 GMEP (2 Node Mix) & $\mathcal{O}(n^3L + n|\mathcal{A}|L+ 2n^2ML + n^2M^2L  )$\\
 \hline
\end{tabular}
\label{Tab:complex}
\end{table}

\section{Performance Analysis}
In this section, we examine the performance of the GMEP
algorithm for MIMO detection in high-order high-dimensional
scenarios. We have used a MATLAB based simulation and averaged our results for $10^5$ symbols. We consider two scenarios of increasing dimension: $n = m = 8$ and $n = m = 12$. We specifically look at scenarios with $n=m$ since LMMSE does not perform well there. The detector performance is shown in
terms of the Symbol Error Eate (SER) as a function of the SNR. In both scenarios, we compare the performance of GMEP, EP, Zero Forcing (ZF) \cite{tse2004fundamentals} and LMMSE \cite{tse2004fundamentals}. 

Also, we have opted to reduce the number of mixture components using:
\[
    \alpha_k = 
\begin{cases}
    1   ,& q^{(l)}_{-s}\left(u_s=k\right) > 10^{-3}\\
    0,              & \text{otherwise}
\end{cases}
\]
This choice has empirically provided reasonable results and reduced the mixture order significantly (for 256-QAM, approximately 2 components in a mixture for SER $=2\cdot 10^{-3}$). For the GMM message smoothing we have used $\beta = 0.8$ for $L=2$ and $\beta=1$ for $L=1$. Also, we have allowed two messages to be GMMs when needed (if more than 1 prior update is negative).

\begin{figure}
    \centering
    \includegraphics[width=1.\linewidth]{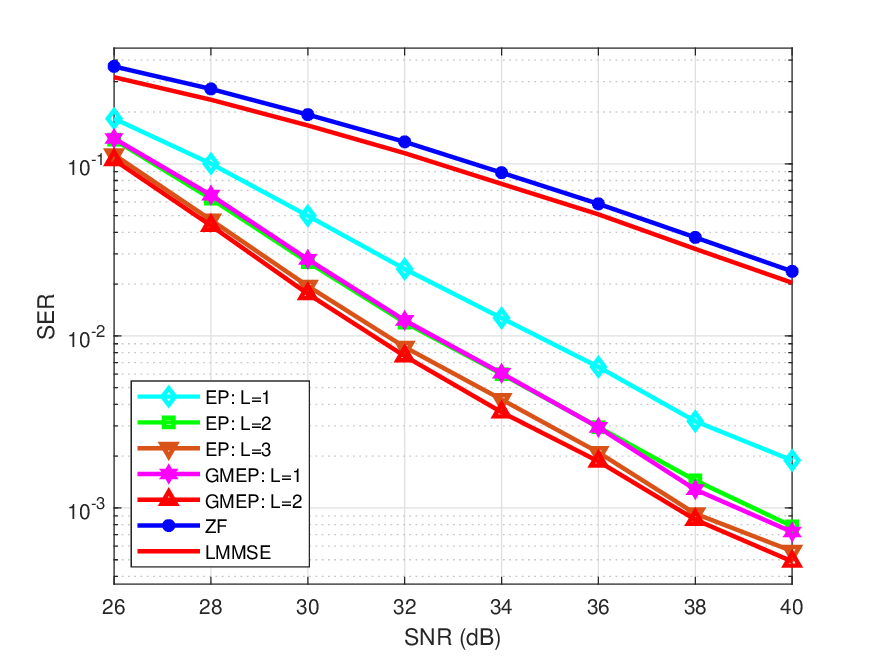}
    \caption{Comparison of various detectors in a 8x8 system, 64-QAM}
    \label{fig:8x8}
\end{figure}

We first consider $n = m = 8$ for a 64-QAM constellation. In Fig. \ref{fig:8x8}, we can observe that GMEP with $L=2$ and $L=1$ outperform all the other algorithms including EP with $L=2$. For $L=1$, GMEP has a $2.5$dB gain over EP with $L=1$ and for $L=2$, GMEP has a gain of $1.5$dB over EP with $L=2$. We also note that GMEP with $L=2$ is better than EP with $L=3$ which has higher complexity than GMEP. Both EP and GMEP significantly outperform LMMSE and ZF as expected.

\begin{figure}
    \centering
    \includegraphics[width=1.\linewidth]{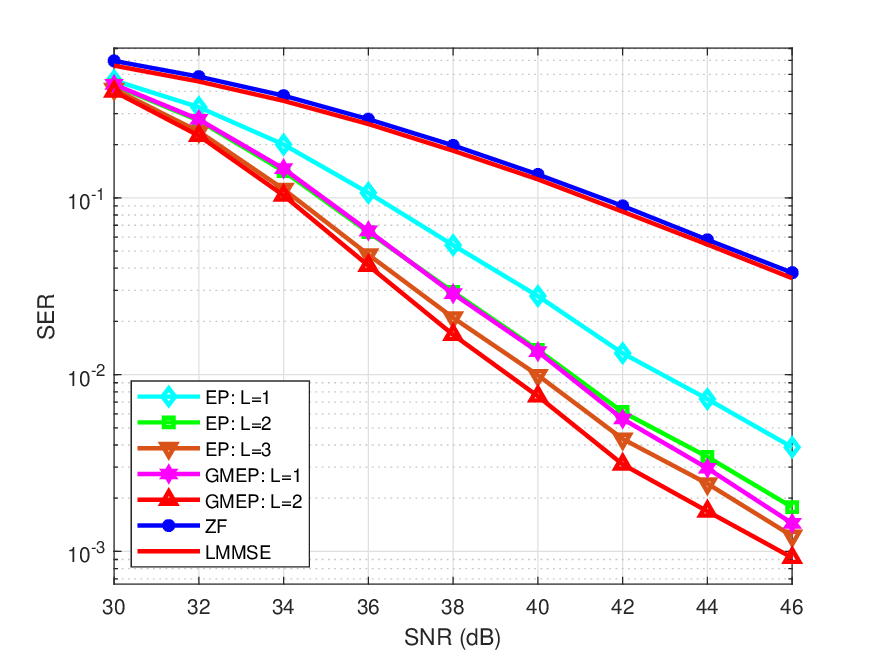}
    \caption{Comparison of various detectors in a 12x12 system, 256-QAM}
    \label{fig:12x12}
\end{figure}

\begin{figure}
    \centering
    \includegraphics[width=1.\linewidth]{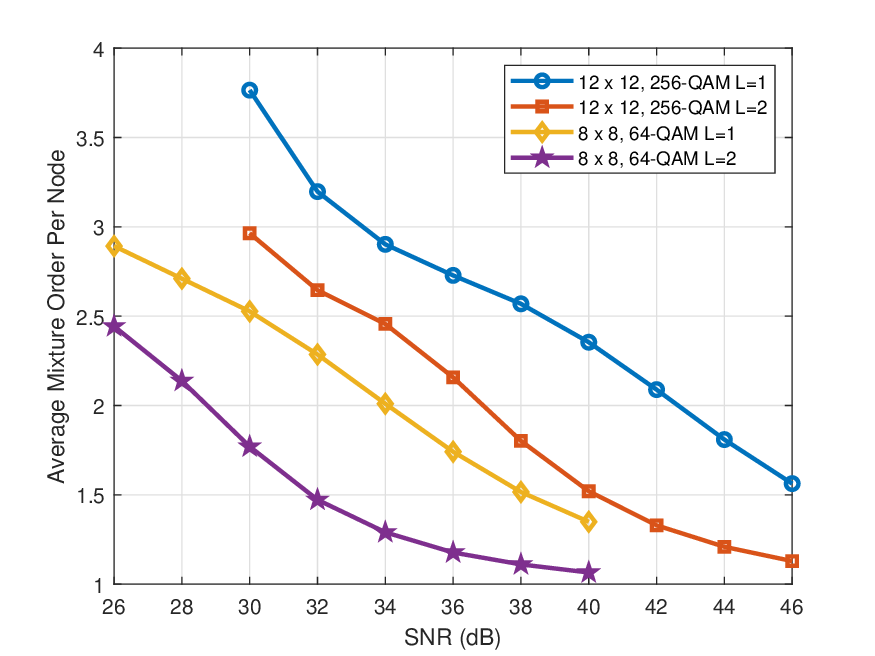}
    \caption{Average mixture order ($M$) per variable node}
    \label{fig:complexity}
\end{figure}

Next, we consider a scenario with $n = m = 12$ with 256-QAM. In Fig. \ref{fig:12x12}, we can see that GMEP with $L=1$ has comparable performance to EP with $L=2$ but since $n > M^2$  for high SNR as shown in Fig.\ref{fig:complexity}, the complexity of GMEP is lower and thus favorable. The same happens for EP with $L=3$ and GMEP with $L=2$, but the gap is larger than for the $n-m-8$ case. Moreover, for $L=1$, GMEP is $3$dB better than EP and for $L=2$, GMEP is approximately $2$dB better than EP.

We have analyzed the expected mixture order for a variable node in Fig.\ref{fig:complexity}. We can see that the expected mixture order decreases as the SNR increases and as the iteration increases, as expected. It is also clear that for the same SNR, the expected mixture order is larger for a larger constellation size. The complexity of GMEP is dominated by $\mathcal{O}(n^2M^2)$ as opposed to EP's $\mathcal{O}(n^3)$ as detailed in Table \ref{Tab:complex}. Therefore, GMEP with $L$ iterations has lower complexity than EP with $L+1$ iterations (for sufficiently $M^2 < n$). In terms of SER performance, it can be seen in Fig.\ref{fig:8x8} and Fig.\ref{fig:12x12}, that GMEP $L=1$ and EP $L=2$ have similar SER, thus GMEP is favorable. 

\section{Discussion}
In this paper a MIMO detection algorithm based on a GMM approximation for EP messages was proposed. The manipulation of the GMM messages is performed with a particular focus on balancing the trade-off between computational complexity and detection performance. Through empirical simulations, we demonstrate the efficacy of our proposed scheme in achieving superior detection performance while maintaining practical feasibility for real-time implementation. Our findings highlight the potential of advanced MIMO detection algorithms to significantly enhance the performance of future wireless communication systems, especially for large array order and symbol constellation MIMO systems. In addition, our algorithm provides soft information on the symbols thus enabling seamless integration to a coded communications system.

\bibliographystyle{IEEEtran}
\bibliography{mybib}

\begin{thebibliography}{10}
\providecommand{\url}[1]{#1}
\csname url@samestyle\endcsname
\providecommand{\newblock}{\relax}
\providecommand{\bibinfo}[2]{#2}
\providecommand{\BIBentrySTDinterwordspacing}{\spaceskip=0pt\relax}
\providecommand{\BIBentryALTinterwordstretchfactor}{4}
\providecommand{\BIBentryALTinterwordspacing}{\spaceskip=\fontdimen2\font plus
\BIBentryALTinterwordstretchfactor\fontdimen3\font minus \fontdimen4\font\relax}
\providecommand{\BIBforeignlanguage}[2]{{%
\expandafter\ifx\csname l@#1\endcsname\relax
\typeout{** WARNING: IEEEtran.bst: No hyphenation pattern has been}%
\typeout{** loaded for the language `#1'. Using the pattern for}%
\typeout{** the default language instead.}%
\else
\language=\csname l@#1\endcsname
\fi
#2}}
\providecommand{\BIBdecl}{\relax}
\BIBdecl

\bibitem{goldsmith2003capacity}
A.~Goldsmith, S.~A. Jafar, N.~Jindal, and S.~Vishwanath, ``Capacity limits of mimo channels,'' \emph{IEEE Journal on selected areas in Communications}, vol.~21, no.~5, pp. 684--702, 2003.

\bibitem{zheng2003diversity}
L.~Zheng and D.~N.~C. Tse, ``Diversity and multiplexing: A fundamental tradeoff in multiple-antenna channels,'' \emph{IEEE Transactions on information theory}, vol.~49, no.~5, pp. 1073--1096, 2003.

\bibitem{he2020model}
H.~He, C.-K. Wen, S.~Jin, and G.~Y. Li, ``Model-driven deep learning for mimo detection,'' \emph{IEEE Transactions on Signal Processing}, vol.~68, pp. 1702--1715, 2020.

\bibitem{boutros2003soft}
J.~Boutros, N.~Gresset, L.~Brunel, and M.~Fossorier, ``Soft-input soft-output lattice sphere decoder for linear channels,'' in \emph{GLOBECOM'03. IEEE Global Telecommunications Conference (IEEE Cat. No. 03CH37489)}, vol.~3.\hskip 1em plus 0.5em minus 0.4em\relax IEEE, 2003, pp. 1583--1587.

\bibitem{burg2005vlsi}
A.~Burg, M.~Borgmann, M.~Wenk, M.~Zellweger, W.~Fichtner, and H.~Bolcskei, ``Vlsi implementation of mimo detection using the sphere decoding algorithm,'' \emph{IEEE Journal of solid-state circuits}, vol.~40, no.~7, pp. 1566--1577, 2005.

\bibitem{goldberger2011mimo}
J.~Goldberger and A.~Leshem, ``Mimo detection for high-order qam based on a gaussian tree approximation,'' \emph{IEEE transactions on information theory}, vol.~57, no.~8, pp. 4973--4982, 2011.

\bibitem{goldberger2013improved}
J.~Goldberger, ``Improved mimo detection based on successive tree approximations,'' in \emph{2013 IEEE International Symposium on Information Theory}.\hskip 1em plus 0.5em minus 0.4em\relax IEEE, 2013, pp. 2004--2008.

\bibitem{guo2006algorithm}
Z.~Guo and P.~Nilsson, ``Algorithm and implementation of the k-best sphere decoding for mimo detection,'' \emph{IEEE Journal on selected areas in communications}, vol.~24, no.~3, pp. 491--503, 2006.

\bibitem{liu2008modified}
T.-H. Liu and Y.-L.~Y. Liu, ``Modified fast recursive algorithm for efficient mmse-sic detection of the v-blast system,'' \emph{IEEE Transactions on Wireless Communications}, vol.~7, no.~10, pp. 3713--3717, 2008.

\bibitem{studer2008soft}
C.~Studer, A.~Burg, and H.~Bolcskei, ``Soft-output sphere decoding: Algorithms and vlsi implementation,'' \emph{IEEE Journal on selected areas in Communications}, vol.~26, no.~2, pp. 290--300, 2008.

\bibitem{vsvavc2013soft}
P.~{\v{S}}va{\v{c}}, F.~Meyer, E.~Riegler, and F.~Hlawatsch, ``Soft-heuristic detectors for large mimo systems,'' \emph{IEEE Transactions on Signal Processing}, vol.~61, no.~18, pp. 4573--4586, 2013.

\bibitem{yedidia2003understanding}
J.~S. Yedidia, W.~T. Freeman, Y.~Weiss \emph{et~al.}, ``Understanding belief propagation and its generalizations,'' \emph{Exploring artificial intelligence in the new millennium}, vol.~8, no. 236--239, pp. 0018--9448, 2003.

\bibitem{cespedes2014expectation}
J.~Cespedes, P.~M. Olmos, M.~S{\'a}nchez-Fern{\'a}ndez, and F.~Perez-Cruz, ``Expectation propagation detection for high-order high-dimensional mimo systems,'' \emph{IEEE Transactions on Communications}, vol.~62, no.~8, pp. 2840--2849, 2014.

\bibitem{jeon2015optimality}
C.~Jeon, R.~Ghods, A.~Maleki, and C.~Studer, ``Optimality of large mimo detection via approximate message passing,'' in \emph{2015 IEEE International Symposium on Information Theory (ISIT)}.\hskip 1em plus 0.5em minus 0.4em\relax IEEE, 2015, pp. 1227--1231.

\bibitem{wu2014low}
S.~Wu, L.~Kuang, Z.~Ni, J.~Lu, D.~Huang, and Q.~Guo, ``Low-complexity iterative detection for large-scale multiuser mimo-ofdm systems using approximate message passing,'' \emph{IEEE Journal of Selected Topics in Signal Processing}, vol.~8, no.~5, pp. 902--915, 2014.

\bibitem{minka2013expectation}
T.~P. Minka, ``Expectation propagation for approximate bayesian inference,'' \emph{arXiv preprint arXiv:1301.2294}, 2013.

\bibitem{loeliger2007factor}
H.-A. Loeliger, J.~Dauwels, J.~Hu, S.~Korl, L.~Ping, and F.~R. Kschischang, ``The factor graph approach to model-based signal processing,'' \emph{Proceedings of the IEEE}, vol.~95, no.~6, pp. 1295--1322, 2007.

\bibitem{uchoa2015iterative}
A.~G. Uchoa, C.~T. Healy, and R.~C. de~Lamare, ``Iterative detection and decoding algorithms for mimo systems in block-fading channels using ldpc codes,'' \emph{IEEE Transactions on Vehicular Technology}, vol.~65, no.~4, pp. 2735--2741, 2015.

\bibitem{woodbury1950inverting}
M.~A. Woodbury, \emph{Inverting modified matrices}.\hskip 1em plus 0.5em minus 0.4em\relax Department of Statistics, Princeton University, 1950.

\bibitem{tse2004fundamentals}
D.~Tse and P.~Viswanath, ``Fundamentals of wireless communication12,'' \emph{Notes}, p. 583, 2004.

\end{thebibliography}

\end{document}